\documentclass[useAMS,usenatbib,usegraphicx]{mn2e}
\usepackage{graphicx,times,epsfig,rotating}

\def\aap{A\&A}

\def\apjl{ApJ}
\def\apjs{ApJS}

\newcommand{\kmsend}{\mbox{km s$^{-1}$}}

\newcommand{\msun}{\mbox{M$_{\sun}$ }}
\newcommand{\msunend}{\mbox{M$_{\sun}$}}

\newcommand{\lco}{\mbox{L$_{\rm CO}$}}

\newcommand{\cmthree}{\mbox{cm$^{-3}$}}

\newcommand{\htwo}{\mbox{H$_2$}}

\newcommand{\z}{\mbox{$z$}}

\newcommand{\sunrise}{\mbox{\sc sunrise}}
\newcommand{\gadget}{\mbox{\sc gadget-3}}
\newcommand{\starburst}{\mbox{\sc starburst99}}

\newcommand{\turtlebeach}{\mbox{\sc turtlebeach}}

\newcommand{\xco}{\mbox{$X_{\rm CO}$}}
\newcommand{\xcounits}{\mbox{cm$^{-2}$/K-km s$^{-1}$}}

\newcommand{\alphaco}{\mbox{$\alpha_{\rm CO}$}}
\newcommand{\alphacounits}{\mbox{\msun pc$^{-2}$ (K-\kmsend)$^{-1}$}}

\newcommand{\msunpcsq}{\mbox{$\msun {\rm pc}^{-2}$}}

\title[The Constant CO-\htwo \ Conversion Factor in the Milky Way]{Why
  is the Milky Way $X$-factor Constant?}

\author[Narayanan \& Hopkins]{Desika\,
  Narayanan$^{1}$\thanks{E-mail:
    dnarayanan@as.arizona.edu}\thanks{Bart J Bok Fellow} \& Philip
  F. Hopkins$^{2}$\thanks{Einstein Fellow}\\$^{1}$Steward Observatory,
  University of Arizona, 933 N Cherry Ave, Tucson, Az, 85721\\$^{2}$Department of Astronomy and Theoretical Astrophysics Center, University of California, Berkeley, Berkeley, CA, 94720}
\begin{document}

\date{Submitted to MNRAS Letters}

\pagerange{\pageref{firstpage}--\pageref{lastpage}} \pubyear{2010}

\maketitle

\label{firstpage}

\begin{abstract}
The CO-\htwo \ conversion factor (\xco; otherwise known as the
$X$-factor) is observed to be remarkably constant in the Milky Way and
in the Local Group (aside from the SMC).  To date, our understanding
of why \xco \ should be so constant remains poor.  Using a combination
of extremely high resolution ($\sim 1 $ pc) galaxy evolution
simulations and molecular line radiative transfer calculations, we
suggest that \xco \ displays a narrow range of values in the Galaxy
due to the fact that molecular clouds share very similar physical
properties. In our models, this is itself a consequence of stellar
feedback competing against gravitational collapse.  GMCs whose
lifetimes are regulated by radiative feedback show a narrow range of
surface densities, temperatures and velocity dispersions with values
comparable to those seen in the Milky Way.  As a result, the
$X$-factors from these clouds show reasonable correspondence with
observed data from the Local Group, and a relatively narrow range.  On
the other hand, feedback-free clouds collapse to surface densities
that are larger than those seen in the Galaxy, and hence result in
$X$-factors that are systematically too large compared to the Milky
Way's. We conclude that radiative feedback within GMCs can generate
cloud properties similar to those observed in the Galaxy, and hence a
roughly constant Milky Way X-factor in normal, quiescent clouds.

\end{abstract}


\section{Introduction}
\label{section:introduction}

Historically, determining \htwo \ gas masses in the interstellar
medium (ISM) of galaxies has relied on the usage of tracer molecules.
This owes to the fact that \htwo \ requires temperatures $\sim 500$ K
to excite the rotational lines, and is thus a poor tracer of cold
($\sim 10-100 $K) giant molecular clouds (GMCs). Carbon Monoxide
($^{12}$CO; hereafter, CO) is the second most abundant molecule, and
has strong lines in a readily accessible window of atmospheric
transmission.  For this reason, CO is the most commonly employed
tracer of \htwo.  However, utilising CO as a measure of \htwo \ does
not come without uncertainty.  

At the heart of converting CO line fluxes to \htwo \ masses is the
CO-\htwo \ conversion factor.  The conversion factor relates either
\htwo \ column densities to velocity-integrated CO line intensity
(\xco; alternatively, the $X$-factor), or \htwo \ gas mass to CO line
luminosity (\alphaco).  The two are related via \xco \ (\xcounits) =
$6.3 \times 10^{19} \times \alphaco$ \ (\alphacounits).  Uncertainties
in CO abundances, \htwo \ gas fractions, and radiative transfer all
complicate our understanding of \xco.

In principle, \xco \ can be empirically calibrated with an independent
measure of \htwo \ gas masses.  Efforts along these lines have used CO
line widths (combined with an assumption regarding the dynamical state
of the GMC), dust mass measurements (combined with an assumed
dust-to-gas ratio), or gamma-ray observations from GMCs to determine
the CO-\htwo \ conversion factor
\citep{lar81,sol87,dic75,dev87,pin08,ler11,mag11,blo86,str96,abd10b,del11}.
These measurements have all arrived at the conclusion that the
$X$-factor in Milky Way GMCs is remarkably constant, displaying a
relatively narrow range of $\xco \approx 2-4\times10^{20} \xcounits$
($\alphaco \approx 3-6 \ \alphacounits$).  Beyond this, observations
of GMCs in relatively normal galaxies within the Local Group
\citep[that is, excluding the Small Magallenic Cloud;][]{ler11} have
evidenced similar $X$-factors as in the Galaxy.  

This said, not all galaxies exhibit $X$-factors comparable to the
relatively constant Local Group disc galaxy value.  In particular,
heavily star-forming systems at low and high-\z \ appear to have
$X$-factors roughly a factor 2-10 lower than the Milky Way average
\citep[e.g. ][]{dow98,tac08,mei10,nar11d}, whereas low-metallicity
galaxies can have $X$-factors up to a factor $\sim 100$ higher than
the Galactic mean
\citep[e.g. ][]{wil95,ari96,isr97,bos02,bol08,ler11,gen12,sch12}.
Critical questions include: (1) what is the origin of the
nearly-constant $X$-factor in nearby disks, (2) why is \xco
\ depressed in high gas-surface density environments, and (3) why is
the conversion factor elevated in low metallicity galaxies?

In recent years, there has been a flurry of interest from theorists in
understanding \xco \ on scales ranging from GMCs to cosmological
simulations of galaxy formation.  Models have made great headway in
understanding the latter two questions.  MHD calculations of evolving
GMCs \citep{she11a,she11b}, cosmological galaxy formation calculations
\citep{fel12a,lag12} and hydrodynamic galaxy evolution calculations
coupled with radiative transfer \citet{nar11b,nar12a} have all
converged on a picture in which low-metallicity galaxies have large
fractions of CO-dark molecular gas (due to photodissociation of CO in
regions of low dust extinction\footnote{This is the presumed origin
  for the large observed $X$-factors in the SMC \citep{ler11}.}), and
hence large \xco.  Similarly, galaxy merger models by \citet{nar11b}
and \citet{nar12a} have shown that starburst environments can force
large gas temperatures and velocity dispersions which increase the CO
intensity at a given \htwo \ gas mass, thus reducing \xco.

This said, thus far no model that considers a full ensemble of clouds
on galaxy-wide scales has been able to explain {\it why} the
$X$-factor has such a narrow range of values in the Milky Way and
nearby galaxies.  To understand this requires knowledge of the
physical state of GMCs on highly-resolved ($\sim$ pc) scales, but
sampled over the scales of entire galaxies.  That is, one ideally
should be able to super-resolve GMCs while capturing the effect of the
larger galactic environment on cloud evolution\footnote{In
  \citet{nar11b}, we implemented a subresolution model for the
  velocity dispersions and surface densities of GMCs in idealised
  galaxy evolution models.  While these models suggested that the
  origin of a roughly constant Galactic $X$-factor owed to a
  relatively narrow range of physical properties in the model GMCs,
  this conclusion was not entirely independent of subresolution
  assumptions.}. Moreover, without explicit models for feedback, GMCs
will inevitably collapse without limit (to arbitrarily high densities)
and turn most of their mass into stars. This is in stark disagreement
with observations indicating inefficient star formation and relatively
short GMC lifetimes \citep[e.g. ][]{eva99,kru07b,shi07,ken12}.

 Recently, \citet{hop11b} have implemented various forms of stellar
 feedback into idealised galaxy evolution simulations that have
 allowed for $\sim$ pc-scale resolution on galaxy-wide scales.  These
 simulations move beyond previous models that employ subresolution
 assumptions governing molecular cloud evolution
 \citep[e.g.][]{nar11b,nar12a} and allow us to super-resolve GMCs on
 galaxy-wide scales.  These simulations have been utilised by
 \citet{hop12} to show that a model in which radiative feedback from
 massive stars dominates the life-cycle of molecular clouds
 successfully reproduces many observed physical properties and scaling
 relations (e.g. ``Larson's Laws'') of GMCs, including their surface
 densities, velocity dispersions, size distributions and mass spectra.
 Here, we employ these simulations to model a Milky Way-like disc
 galaxy and ask whether GMCs whose physical properties are governed by
 radiative feedback can explain $X$-factor properties (i.e. their
 constancy) of observed Local Group GMCs.

\section{Methods}

We simulate the hydrodynamic evolution of a Milky Way-like galaxy with
a substantially modified version of \gadget, a smoothed-particle
hydrodynamics (SPH) code \citep{spr05a,spr05b}.  The main code
modifications and model details are described in \citet{hop11b,hop12}
and we describe only the important aspects here.

We initialise an exponential disc according to the \citet{mo98} model
within a live dark matter halo of mass $1.6 \times 10^{12} \msun$
\ with a \citet{her90} density profile and concentration parameter
$c=12$.  The baryon, bulge, disc and gas masses are initialised at
($M_{\rm bar},m_b,m_d,m_g$) = ($7.1,1.5,4.7,0.9$)$\times 10^{10}
\msunend$, and scale-lengths ($h_d,h_g,z_0$)=$(3.0,6.0,0.3$) kpc.  The
simulations are run with $\sim3\times 10^{7}$ hydrodynamic particles,
with a smoothing length of $\sim 1$pc.

For the purposes of the SPH simulations, the gas is allowed to cool to
$\sim 100$ K (though see below regarding further refinements on this
temperature structure in post-processing), and the HI-\htwo \ balance
is determined following the semi-analytic model of
\citet{kru08,kru09a}.  Stars form exclusively in \htwo \ gas that is
self-gravitating on the smallest resolved scales.  Stars form at an
instantaneous rate of $\dot{\rho}_* = \rho/t_{\rm ff}(\rho)$; because
feedback can prevent further star formation once stars form, the
average efficiency of star formation in dense gas is typically $\sim
1\%$. 

The most important aspect of the models is a range of mechanisms by
which stellar feedback can impact the ISM.  Throughout, we assume a
\citet{kro02} stellar IMF, and utilise \starburst \ \citep{lei99} for
all stellar luminosity, mass return and supernova rate calculations as
a function of stellar age and metallicity.  The mechanisms of stellar
feedback include:

\begin{enumerate}
\item {\bf Local Momentum Deposition}: For the purposes of this work,
  the most important source of feedback is local momentum deposition
  by stellar radiation, mass return from stellar winds and
  supernova. At each timestep, the nearest density peak to a given gas
  particle is determined to represent a clump inside of a GMC.  The
  total stellar radiation from all star particles inside a sphere
  defined by the distance between the gas particle and density peak
  are summed; the radiation from these stars is then used to
  determine the momentum flux.

The momentum flux from radiation is given by $\dot{P}_{\rm rad} =
(1+\tau_{\rm IR}) \times (L/c)$, where $\tau_{\rm IR} = \Sigma_{\rm
  gas} \kappa_{\rm IR}$.  $\Sigma_{\rm gas}$ is the gas surface
density calculated directly from the simulation, and $\kappa_{\rm IR}$
is approximated by $\kappa_{\rm IR} = 5(Z/Z_\odot)$g$^{-1}$ cm$^2$.  

We also include direct momentum injection from SNe and stellar winds,
whose momentum deposition rates are directly calculated from
\starburst \ and injected to the gas within a smoothing length of the
star particle.  This source of feedback is typically subdominant
compared to radiation pressure in galaxies as massive as a MW analog.

\item {\bf Thermal Energy Input from Supernova and Stellar Wind
  Shock-Heating}: For the supernovae (SNe), we tabulate Type I and II
  SNe rates from \citet{man06} and \starburst, respectively, for all
  star particles and determine from a stochastic procedure if a SN
  occurred at each timestep.  When a SN occurs, thermal energy is
  injected into the gas within a smoothing length of the star
  particle. For stellar winds, we inject the tabulated mechanical
  power as a function of stellar age and metallicity to the gas within
  a smoothing length of the star.

\item {\bf Photo-Heating of HII Regions} The production rates of
  ionising photons from star particles is calculated and used to
  determine the extent of HII regions surrounding stars (allowing appropriately for 
  overlapping regions).  The
  temperatures of these HII regions are heated to $10^4$ K if the gas falls below that threshold.

\item {\bf Long-Range Radiation Pressure} Photons that escape the
  local GMC (not accounted for in {\bf(i)}), after being appropriately
  attenuated/absorbed, are propagated to large distances along direct
  rays, where local absorption is calculated by integrating over a
  frequency-dependent opacity (which also scales linearly with
  metallicity). The appropriate radiation pressure forces are then
  imparted.

\end{enumerate}

Predicting the molecular emission requires calculating the correct
temperature (including full radiative transfer effects) of molecular
gas below $T=100\,$K (the approximate floor imposed by the cooling
tables in the simulations). This is prohibitively expensive to perform
on-the-fly, so in this paper we do so in post-processing. But we do
not expect changes in the thermal pressure at such low temperatures to
have any dynamical effect on the simulations. We calculate the
temperature utilising the methodology described in \citet{gol01} and
\citet{kru11a}.  The dominant heating processes of the \htwo \ gas are
cosmic ray heating, the grain photoelectric effect, and dust-gas
thermal exchange.  The dominant cooling terms are cooling by CO, CII
and dust.  If the gas and dust are in thermal balance, then we have
the following equations (where $\Gamma$ represents heating terms, and
$\Lambda$ represents cooling terms):

\begin{eqnarray}
\Gamma_{\rm pe} + \Gamma_{\rm CR} - \Lambda_{\rm line} + \Psi_{\rm gd} = 0\\
\Gamma_{\rm dust} - \Lambda_{\rm dust} - \Psi_{\rm gd} = 0  
\end{eqnarray}
The equation is solved by simultaneously iterating on the temperatures
of the gas and dust

We assume a Galactic cosmic ray heating rate and a grain photoelectric
effect proportional to the local FUV intensity.  We refer the reader
to \citet{kru11c} for the specific values employed in the model.  In
short, however, the temperature can be thought of as
density-dependent.  At high densities ($n>10^4 \cmthree$), dust and
gas exchange energy efficiently, and the gas temperature rises to the
dust temperature.  At low densities ($n<10^2 \cmthree$), cosmic rays
dominate the heating. For a Milky Way cosmic ray flux, this
corresponds to GMCs with temperatures $\sim 8-10 $ K.  Intermediate
densities have temperatures in between the dust temperature and cosmic
ray-determined temperature.

The dust temperature is calculated utilising \sunrise, a publicly
available Monte Carlo dust radiative transfer code \citep[see][for
  code descriptions, as well as \citet{hay11,hay12a,hay12b} for
  further details]{jon10a,jon10b}.  We utilise the simulation set-up
described in \citet{nar11b}, and refer the reader there for more
details.  In practice, the bulk of the gas remains below density
$n=10^4 \cmthree$, the density at which dust-gas energy exchange
becomes efficient.

The line cooling term is calculated in each cell via a 1D escape
probability code \citep{kru07}.  We assume a fractional carbon
abundance of $1.5\times10^{-4} Z'$, where $Z'$ is the metallicity with
respect to solar.  The fraction of hydrogen where the carbon is in the
form of CO is well approximated from both semi-analytic \citep{wol10}
and numerical models \citep{glo11}:
\begin{equation}
\label{eq:abundance}
f_{\rm CO} = f_{\rm H2} \times e^{-4(0.53-0.045 {\rm
    ln}\frac{G_0'}{n_{\rm H}/{\rm cm^{-3}}}-0.097 {\rm ln}Z')/A_{\rm v}}
\end{equation}
where $G_0'$ is the FUV intensity relative to the Solar neighbourhood.
Physically, Equation~\ref{eq:abundance} describes the
photodissociation of CO from UV photons, and its ability to survive
behind sufficient columns of dust.  When $f_{\rm CO} > 0.5$, we assume
that CO dominates the line cooling; else, CII.

Finally, we note that we do not explicitly include the effects of
heating by turbulent dissipation.  In \citet{nar11b}, we performed
tests in which the contribution of viscous dissipation and adiabatic
compression to the turbulent heating rate were included in the model.
These tests showed that, for quiescent discs, this source of heating
only affects the final temperature by a few percent.

With the physical and chemical state of the molecular gas known, we
utilize \turtlebeach, a 3D non-local thermodynamic equilibrium
adaptive mesh Monte Carlo line radiative transfer code to calculate
the velocity integrated CO line intensity
\citep{nar06b,nar08a,nar11a,nar11b}.  We refer the reader to \citet{nar11b}
for the formal equations, and only summarise the relevant points here.

CO line emission is set by the level populations.  The source function
for a given transition $u\rightarrow l$ is given by:
\begin{equation}
S_{\nu}=\frac{n_{u}A_{ul}}{(n_{l}B_{lu}-n_{u}B_{ul})}
\end{equation}
where $A_{ul}, B_{lu}$ and $B_{ul}$ are the Einstein rate
coefficients, and $n$ are the level populations.  

We first calculate the level populations within a given cell utilising
the escape probability formalism \citep{kru07}.  We emit model photons
from each cell isotropically with emission frequency drawn from a
Gaussian profile function.  When the photon passes through a cell, it
sees an opacity of:
\begin{equation}
\alpha_\nu^{ul}(\rm gas)= \frac{h
    \nu_{ul}}{4\pi}\phi(\nu)(\it n_lB_{lu}-n_uB_{ul})
\end{equation}
where $\nu$ is the transition frequency, and $\phi(\nu)$ is the line
profile function that takes into account the effects of line of sight
velocity offsets in the opacity.

Once the model photons have all been emitted, the level populations
are updated by assuming detailed balance \citep{nar11b}.  The
collisional rate coefficients are taken from the {\it Leiden Atomic
  and Molecular Database} \citep{sch05}.  This process is iterated
upon until the level populations are converged to $<1\%$ across all
cells.

GMCs within the model are identified via a friends of friends finder
with a linking length of $20\%$ of the mean cell size.  Tests have
shown that the results are not substantially sensitive to this choice. 

\section{Why is the X-factor Constant?}

To first order, \xco \ can be thought of as the column density of GMCs
divided by the product of their temperature and velocity dispersion.
Formally, \xco = $N_{\rm H2}/W_{\rm CO}$, where $W_{\rm CO}$ is the
velocity-integrated CO intensity.  When the gas is in local
thermodynamic equilibrium (as CO J=1-0 almost always is), the
amplitude of the emission line is proportional to the gas kinetic
temperature $T_{\rm K}$.  Similarly, because CO (J=1-0) is typically
optically thick within GMCs, increasing the velocity dispersion of the
gas increases the emergent CO intensity.  As a result, $W_{\rm CO}$
increases for both increasing kinetic temperature, as well as
increasing velocity dispersion.  So, to ask why the $X$-factor in MW
GMCs is nearly constant is to ask why observed gas temperatures,
velocity dispersions, and surface densities have a narrow distribution
of values.

\subsection{The Physical Properties of GMCs in Galaxy Discs}

\begin{figure*}
\hspace{-1cm}
\includegraphics[angle=90,scale=0.7]{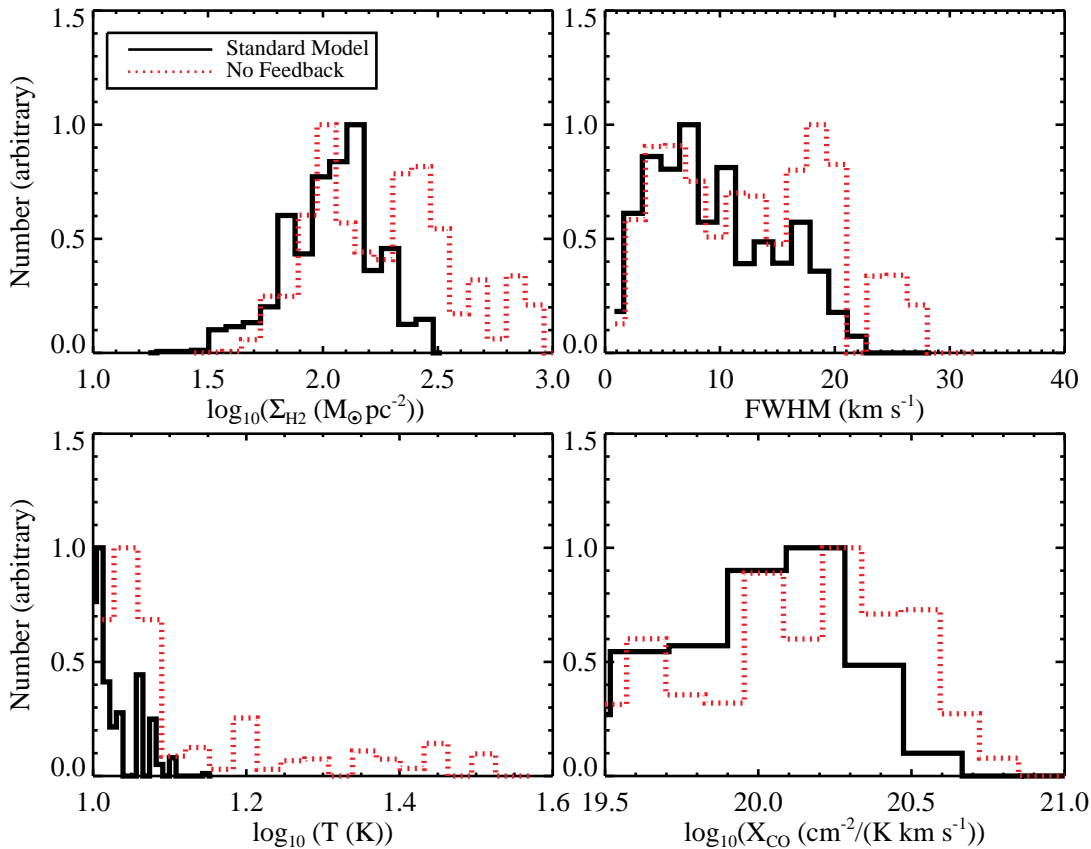}
\caption{Mass-weighted distributions over all GMCs of gas surface
  density, velocity dispersion, temperature, and $X$-factor for our
  fiducial model (solid black line), and a feedback-free model (dotted
  line).  Feedback-free GMCs collapse to higher surface densities than
  models that include feedback, driving more power into large \xco
  \ values.\label{figure:fourplot}}
\end{figure*}

In Figure~\ref{figure:fourplot}, we show the mass-weighted
distributions of GMC temperatures, velocity dispersions, surface
densities, and $X$-factors for our fiducial MW model at a randomly
chosen time snapshot (though the results are consistent for the bulk
of the galaxy's evolution).  As we will see, the physical properties
of the GMCs are generally determined by the radiative feedback that
eventually disrupts the GMC.

In our model disc, the GMCs can be thought of as roughly isothermal,
with temperatures $\sim 10$ K.  The average densities of the GMCs are
relatively low, with $n \approx 10-100 \ \cmthree$.  At these
densities, the gas is energetically decoupled from the dust, and
cosmic rays act as the primary heating source.  For a Galactic cosmic
ray flux (which we assume), this equates to a minimum gas temperature
of roughly $\sim 10$ K.  We note that this is in reasonable agreement
with measurements from the Milky Way.  For example, the mass-weighted
mean temperature for all GMCs in the \citet{sol87} First Quadrant
survey is $\sim 11.3$ K. If we restrict the averaging to only GMCs
with reported temperatures $T > 10 K$ to avoid any potentially
subthermally excited clouds, the mass weighted average temperature is
$\sim 13 $K.  Both values are in good agreement with our modeled
values in Figure~\ref{figure:fourplot}.  

The GMC surface densities display a range of values centred around
$\sim 100 \ \msunpcsq$. We calculate the surface density as
$\Sigma_{\rm H2} = M/(\pi \times R_{\rm C}^2)$, where the radius of
the GMC ($R_{\rm C}$) is half the average of the maximum length in
three orthogonal directions.  While this is method is not free of
geometric effects, it is a reasonable approximation to the methods
used in observations.  These surface densities are comparable to
collapse conditions for gas that forms GMCs.  In a model where the GMC
lifetime is regulated by radiative feedback, GMCs do not collapse
indefinitely.  Once GMCs reach surface densities near 100 \msunpcsq,
feedback from star formation disperses the cloud.  These surface
densities are not far from the average value of the disc, as the model
GMC only lives a few $\times 10^6$ yr (e.g. a few free fall times).
Beyond this, these surface densities are comparable to those seen in
Galactic GMCs, which display a relatively narrow range \citep[][though
  see \citet{lom10}]{lar81,sol87,hey09}.

It is worth noting here that the $X$-factor is implicitly dependent on
the volumetric density of the GMCs residing within a range of $n \sim
50-10^4$ \cmthree.  At larger densities, the gas and dust exchange
energy efficiently, and the gas temperature rises to that of the dust
temperature \citep{gol01,kru11a}, driving it to larger values than the
roughly $\sim 10$ K value seen when cosmic rays dominate the
heating\footnote{Given a Milky Way cosmic ray flux.}.  At lower
densities, the CO may not be in LTE (depending on the degree of line
radiative trapping).  In this regime, the peak CO intensity will no
longer scale with the kinetic temperature of the gas.  As in the case
with the GMC surface densities, radiative feedback suppresses the
formation of excessive amounts of dense gas \citep[][]{hop12d}.  The
median density of GMCs is $\sim 100$ \cmthree, and the distribution
within the galaxy is roughly lognormal in shape.

The GMCs in these simulations are consistent with being marginally gravitationally bound
\citep{hop12}, and have velocity dispersions ranging from a few to
$\sim 20$ \kmsend.  Disruption of the GMC by radiative feedback keeps
the GMC from becoming too strongly self-gravitating, and hence limits
the velocity dispersions. Comparing these to the typical velocity
dispersion of GMCs seen in the Galaxy \citep{sol87}, the range of
modeled GMC velocity dispersions is reasonable.

\subsection{$X$-factor properties in Galaxy discs}

\begin{figure}
\hspace{-1cm}
\includegraphics[angle=90,scale=0.4]{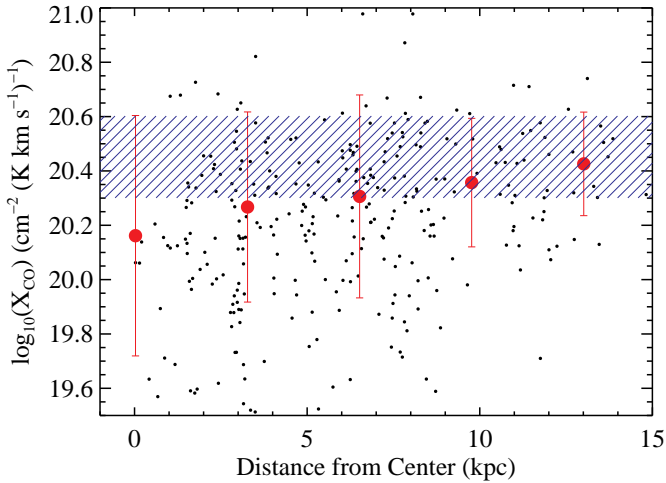}
\caption{$X$-factors \ of model GMCs in our fiducial model galaxy as a
  function of galactocentric distance.  Black circles represent the
  individual GMCs, while the red filled circles denote the median of
  all GMCs within bins of $\sim$ 3 kpc (with the dispersion within the
  bin shown as error bars). The blue shaded region shows the rough
  range of values for the Milky Way. GMCs toward the centre of the
  galaxy tend to have lower $X$-factors due to systematically larger
  velocity dispersions, though the trend is relatively
  weak, and there is significant dispersion.  \label{figure:xco_dist}}
\end{figure}

We can now see why the $X$-factor is relatively constant in Milky
Way-like disc galaxies.  If the evolution of GMCs is largely governed
by radiative feedback, their surface densities and velocity
dispersions display a relatively narrow range of values comparable to
measurements of Galactic GMCs.  If, beyond this, the temperatures of
GMCs are nearly isothermal at $\sim 10$ K, as expected for clouds
where the dominant heating source is cosmic rays with a flux
comparable to the Galaxy's, then the CO-\htwo \ conversion factor will
display a relatively narrow range of values centred around 2-4 $\times
10^{20}$ \ \xcounits.  We see this in the fourth panel of
Figure~\ref{figure:fourplot}.

In Figure~\ref{figure:xco_dist}, we show the relationship between \xco
\ in our model GMCs and their distance from the centre of the galaxy.
The black points denote the individual GMCs, while the red circles and
error bars denote the median and dispersion within roughly 3 kpc
bins. The blue shaded region shows the average Galactic range for the
$X$-factor.  Generally, GMCs toward the centre of the model Milky Way
analog systematically have lower $X$-factors than clouds at larger
distances from the galactic centre. Influenced by a large stellar
potential, as well as other gas, GMCs toward the centre of the galaxy
tend to have larger velocity dispersions than field GMCs, and tend to
be unvirialised.  There is some tentative observational evidence that
\xco \ values in GMCs decrease from the Galactic mean toward the
centre of the Milky Way \citep{oka98,str04}, as well as in other nearby
galaxies \citep{san12}.  This said, there is significant dispersion in
observed trends of \xco \ with galactic radius, and some observations
show nearly no depression at all (less than a factor of 2) toward
galactic nuclei (Donovan Meyer et al. in prep.).

\begin{figure}
\hspace{-0.5cm}
\includegraphics[scale=0.4,angle=90]{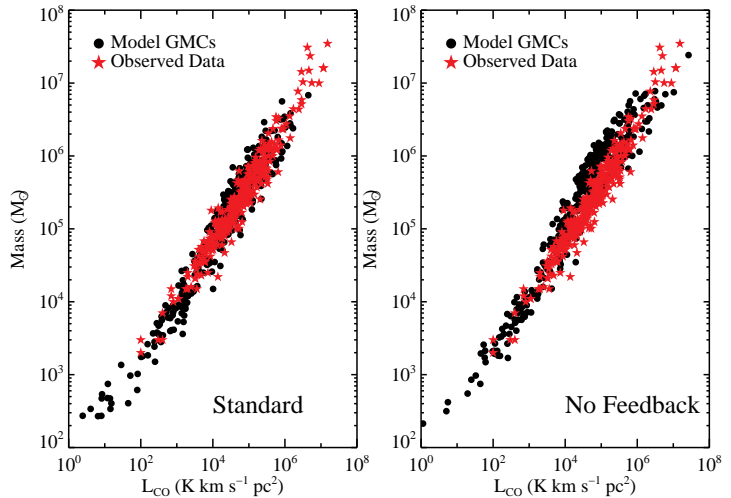}
\caption{Virial mass versus CO luminosity for observed GMCs (red
  stars), and modeled clouds (black filled circles).  Left plot shows
  our fiducial galaxy, and right plot a feedback-free model.  While
  the clouds in our fiducial galaxy show excellent correspondence
  with observed GMCs, the feedback-free clouds systematically has
  $X$-factors that are too large, driving the simulated $M_{\rm
    vir}-L_{\rm CO}$ relation a factor of a few higher than the
  observed points. \label{figure:xco_mlco}}
\end{figure}

\begin{figure}
\hspace{-0.5cm}
\includegraphics[scale=0.4,angle=90]{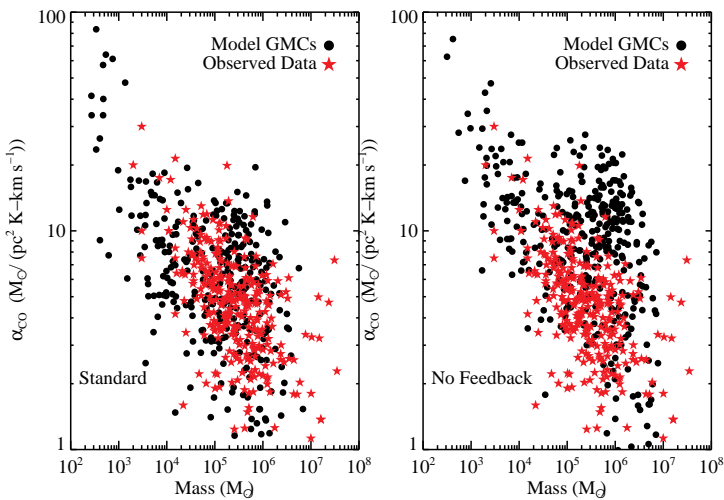}
\caption{\alphaco \ ($M$/\lco) as a function of cloud virial
  mass.  Symbols are as in Figure~\ref{figure:xco_mlco}.  While the
  typical range of modeled CO-\htwo \ conversion factors in our
  fiducial model correspond well with observations, the feedback-free
  model systematically has conversion factors that are too
  large. \label{figure:alpham}}
\end{figure}

The $X$-factor from our model GMCs compares well to the observed
population of GMCs within the Galaxy.  In the left panel of
Figure~\ref{figure:xco_mlco}, we show the virial masses of observed
GMCs in the Galaxy and NGC 6946 against their CO luminosities from
\citet{sol87} and \citet{don12}.  We additionally show the $M_{\rm
  vir}-L_{\rm CO}$ relation for our model galaxies. We defer
discussion of the right panel for \S~\ref{section:feedbackfree}.  The
normalisation of the $M_{\rm vir}-L_{\rm CO}$ relation in the model
GMCs (black solid points), which betrays the conversion of CO to \htwo
\ gas mass, corresponds reasonably well with the observed data (red
stars).  This is shown more explicitly in the left panel of
Figure~\ref{figure:alpham}, which shows the relationship between
\alphaco \ and cloud virial mass for both observed and modeled
GMCs\footnote{We plot in terms of \alphaco \ instead of \xco \ as the
  former is easily calculated from literature measurements of virial
  mass and CO luminosity.  The two forms of the conversion factor are,
  of course, trivally related.}.

\subsection{Feedback-Free Models}
\label{section:feedbackfree}
In order to highlight the role of feedback in setting the physical
properties of the model GMCs, it is worth considering the properties
of a feedback-free model.  At this point, we now highlight the dotted
red line in Figure~\ref{figure:fourplot}, which denotes a model run
with exactly the same initial conditions as our fiducial model, though
with no forms of feedback included. When GMCs first form, their
properties are not especially different in models with and without
feedback, as expected from general models of gravitational instability
and fragmentation \citep{hop12c}. But GMCs in models without feedback
proceed to collapse one-dimensionally, developing a pancake-shaped
geometry and spinning up \citep{hop12}.  While the general surface
density distribution is not terribly dissimilar from our fiducial
model, feedback-free GMCs develop a large tail of very high-density
gas, and show more power toward large $\Sigma_{\rm H2}$.  The velocity
dispersion distribution additionally increases moderately (due to a
spinning up of the contracting GMC), as well as the temperature (due
to more gas at or above the gas-dust coupling density of $\sim 10^4
\cmthree$), but these increases are more modest than the increase in
the GMC surface densities.

The net result of feedback-free GMCs is more power toward large
$X$-factors, and typical $X$-factors a factor $\sim 2-3$ larger than
typical MW GMCs. As an example, in the right panel of
Figure~\ref{figure:xco_mlco}, we plot the $M_{\rm vir}-L_{\rm CO}$
relation for observed GMCs and model GMCs in a feedback-free model.
Compared to our standard model, which shows excellent correspondence
with observed data, the feedback-free model lies a factor of a few
above the $M_{\rm vir}-L_{\rm CO}$ relation.  The right panel of
Figure~\ref{figure:alpham} further empahsises this point by showing
the relationship between modeled conversion factors from feedback-free
GMCs, and observed ones for Galactic clouds.  Beyond this, as noted by
\citet{hop12}, feedback-free models exhibit a host of problems,
including star formation rates well above the observed
Kennicutt-Schmidt relation, gas density distributions (and HCN/CO line
ratios) above those observed \citep{hop12d}, and abnormally low virial
parameters.

\subsection{Summary}
Utilising a combination of high-resolution galaxy evolution
simulations that super-resolve giant molecular clouds and 3D molecular
line radiative transfer calculations, we investigate why the CO-\htwo
\ conversion factor (\xco) is observed to be roughly constant in the
Milky Way and Local Group galaxies (aside from the SMC).

Our main result is that \xco \ is found to be nearly constant because
all GMCs in our model Milky Way have similar physical conditions to
one another.  In particular, \xco \ is determined principally by GMC
surface densities, temperatures, and velocity dispersions.

\begin{itemize}

\item The model GMCs all
have similar surface densities at values near $\sim 100 \ \msunpcsq$.
Models that include radiative feedback limit GMCs from achieving
surface densities much larger than this value due to cloud dispersal.

\item The temperatures of GMCs are dominated by cosmic ray heating.  Given a
Milky Way cosmic ray flux, this results in nearly isothermal GMCs with
temperatures $\sim 8-10 $ K.

\item The GMCs in our model are consistent with being marginally
  bound, resulting in a narrow velocity dispersion range of $5-20$
  \kmsend.
\end{itemize}

These GMCs display a relatively narrow range of physical properties,
and compare well with those observed in the Milky Way.  As a result,
the CO-\htwo \ conversion factor in these clouds additionally shows a
narrow range of values.  Feedback is a necessary element in the model
to control the cloud surface densities, and hence limit the observed
range of \xco.  Feedback-free models collapse to large surface
densities, and hence show excessive power to large \xco \ values.

\section*{Acknowledgements}
 DN thanks Jennifer Donovan Meyer, Lars Hernquist, Mark Krumholz \&
 Eve Ostriker for helpful conversations and acknowledges support from
 the NSF via grant AST-1009452.

\bibliographystyle{apj} \bibliography{/Users/dnarayanan/paper/refs}

\begin{thebibliography}{66}
\expandafter\ifx\csname natexlab\endcsname\relax\def\natexlab#1{#1}\fi

\bibitem[{{Abdo} {et~al.}(2010)}]{abd10b}
{Abdo}, A.~A. {et~al.} 2010, \apjl, 709, L152

\bibitem[{{Arimoto} {et~al.}(1996){Arimoto}, {Sofue}, \& {Tsujimoto}}]{ari96}
{Arimoto}, N., {Sofue}, Y., \& {Tsujimoto}, T. 1996, PASJ, 48, 275

\bibitem[{{Bloemen} {et~al.}(1986){Bloemen}, {Strong}, {Mayer-Hasselwander},
  {Blitz}, {Cohen}, {Dame}, {Grabelsky}, {Thaddeus}, {Hermsen}, \&
  {Lebrun}}]{blo86}
{Bloemen}, J.~B.~G.~M., {Strong}, A.~W., {Mayer-Hasselwander}, H.~A., {Blitz},
  L., {Cohen}, R.~S., {Dame}, T.~M., {Grabelsky}, D.~A., {Thaddeus}, P.,
  {Hermsen}, W., \& {Lebrun}, F. 1986, \aap, 154, 25

\bibitem[{{Bolatto} {et~al.}(2008){Bolatto}, {Leroy}, {Rosolowsky}, {Walter},
  \& {Blitz}}]{bol08}
{Bolatto}, A.~D., {Leroy}, A.~K., {Rosolowsky}, E., {Walter}, F., \& {Blitz},
  L. 2008, \apj, 686, 948

\bibitem[{{Boselli} {et~al.}(2002){Boselli}, {Lequeux}, \& {Gavazzi}}]{bos02}
{Boselli}, A., {Lequeux}, J., \& {Gavazzi}, G. 2002, AP\&SS, 281, 127

\bibitem[{{de Vries} {et~al.}(1987){de Vries}, {Thaddeus}, \&
  {Heithausen}}]{dev87}
{de Vries}, H.~W., {Thaddeus}, P., \& {Heithausen}, A. 1987, \apj, 319, 723

\bibitem[{{Delahaye} {et~al.}(2011){Delahaye}, {Fiasson}, {Pohl}, \&
  {Salati}}]{del11}
{Delahaye}, T., {Fiasson}, A., {Pohl}, M., \& {Salati}, P. 2011, \aap, 531,
  A37+

\bibitem[{{Dickman}(1975)}]{dic75}
{Dickman}, R.~L. 1975, \apj, 202, 50

\bibitem[{{Donovan Meyer} {et~al.}(2012){Donovan Meyer}, {Koda}, {Momose},
  {Fukuhara}, {Mooney}, {Towers}, {Egusa}, {Kennicutt}, {Kuno}, {Carty},
  {Sawada}, \& {Scoville}}]{don12}
{Donovan Meyer}, J., {Koda}, J., {Momose}, R., {Fukuhara}, M., {Mooney}, T.,
  {Towers}, S., {Egusa}, F., {Kennicutt}, R., {Kuno}, N., {Carty}, M.,
  {Sawada}, T., \& {Scoville}, N. 2012, \apj, 744, 42

\bibitem[{{Downes} \& {Solomon}(1998)}]{dow98}
{Downes}, D. \& {Solomon}, P.~M. 1998, \apj, 507, 615

\bibitem[{{Evans}(1999)}]{eva99}
{Evans}, II, N.~J. 1999, \araa, 37, 311

\bibitem[{{Feldmann} {et~al.}(2012){Feldmann}, {Gnedin}, \&
  {Kravtsov}}]{fel12a}
{Feldmann}, R., {Gnedin}, N.~Y., \& {Kravtsov}, A.~V. 2012, \apj, 747, 124

\bibitem[{{Genzel} {et~al.}(2012)}]{gen12}
{Genzel}, R. {et~al.} 2012, \apj, 746, 69

\bibitem[{{Glover} \& {Mac Low}(2011)}]{glo11}
{Glover}, S.~C.~O. \& {Mac Low}, M.-M. 2011, \mnras, 412, 337

\bibitem[{{Goldsmith}(2001)}]{gol01}
{Goldsmith}, P.~F. 2001, \apj, 557, 736

\bibitem[{{Hayward} {et~al.}(2012{\natexlab{a}}){Hayward}, {Jonsson}, {Kere{\v
  s}}, {Magnelli}, {Hernquist}, \& {Cox}}]{hay12a}
{Hayward}, C.~C., {Jonsson}, P., {Kere{\v s}}, D., {Magnelli}, B., {Hernquist},
  L., \& {Cox}, T.~J. 2012{\natexlab{a}}, \mnras, 424, 951

\bibitem[{{Hayward} {et~al.}(2011){Hayward}, {Kere{\v s}}, {Jonsson},
  {Narayanan}, {Cox}, \& {Hernquist}}]{hay11}
{Hayward}, C.~C., {Kere{\v s}}, D., {Jonsson}, P., {Narayanan}, D., {Cox},
  T.~J., \& {Hernquist}, L. 2011, \apj, 743, 159

\bibitem[{{Hayward} {et~al.}(2012{\natexlab{b}}){Hayward}, {Narayanan},
  {Kere{\v s}}, {Jonsson}, {Hopkins}, {Cox}, \& {Hernquist}}]{hay12b}
{Hayward}, C.~C., {Narayanan}, D., {Kere{\v s}}, D., {Jonsson}, P., {Hopkins},
  P.~F., {Cox}, T.~J., \& {Hernquist}, L. 2012{\natexlab{b}}, arXiv/1209.2413

\bibitem[{{Hernquist}(1990)}]{her90}
{Hernquist}, L. 1990, \apj, 356, 359

\bibitem[{{Heyer} {et~al.}(2009){Heyer}, {Krawczyk}, {Duval}, \&
  {Jackson}}]{hey09}
{Heyer}, M., {Krawczyk}, C., {Duval}, J., \& {Jackson}, J.~M. 2009, \apj, 699,
  1092

\bibitem[{{Hopkins}(2011)}]{hop12c}
{Hopkins}, P.~F. 2011, arXiv/1111.2863

\bibitem[{{Hopkins} {et~al.}(2012{\natexlab{a}}){Hopkins}, {Narayanan},
  {Murray}, \& {Quataert}}]{hop12d}
{Hopkins}, P.~F., {Narayanan}, D., {Murray}, N., \& {Quataert}, E.
  2012{\natexlab{a}}, arXiv/1209.0459

\bibitem[{{Hopkins} {et~al.}(2011){Hopkins}, {Quataert}, \& {Murray}}]{hop11b}
{Hopkins}, P.~F., {Quataert}, E., \& {Murray}, N. 2011, \mnras, 417, 950

\bibitem[{{Hopkins} {et~al.}(2012{\natexlab{b}}){Hopkins}, {Quataert}, \&
  {Murray}}]{hop12}
---. 2012{\natexlab{b}}, \mnras, 421, 3488

\bibitem[{{Israel}(1997)}]{isr97}
{Israel}, F.~P. 1997, \aap, 328, 471

\bibitem[{{Jonsson} {et~al.}(2010){Jonsson}, {Groves}, \& {Cox}}]{jon10a}
{Jonsson}, P., {Groves}, B.~A., \& {Cox}, T.~J. 2010, \mnras, 186

\bibitem[{{Jonsson} \& {Primack}(2010)}]{jon10b}
{Jonsson}, P. \& {Primack}, J.~R. 2010, New Astronomy, 15, 509

\bibitem[{{Kennicutt} \& {Evans}(2012)}]{ken12}
{Kennicutt}, Jr., R.~C. \& {Evans}, II, N.~J. 2012, arXiv/1204.3552

\bibitem[{{Kroupa}(2002)}]{kro02}
{Kroupa}, P. 2002, Science, 295, 82

\bibitem[{{Krumholz} {et~al.}(2011{\natexlab{a}}){Krumholz}, {Dekel}, \&
  {McKee}}]{kru11c}
{Krumholz}, M.~R., {Dekel}, A., \& {McKee}, C.~F. 2011{\natexlab{a}},
  arXiv/1109.4150

\bibitem[{{Krumholz} {et~al.}(2011{\natexlab{b}}){Krumholz}, {Leroy}, \&
  {McKee}}]{kru11a}
{Krumholz}, M.~R., {Leroy}, A.~K., \& {McKee}, C.~F. 2011{\natexlab{b}}, \apj,
  731, 25

\bibitem[{{Krumholz} {et~al.}(2008){Krumholz}, {McKee}, \& {Tumlinson}}]{kru08}
{Krumholz}, M.~R., {McKee}, C.~F., \& {Tumlinson}, J. 2008, \apj, 689, 865

\bibitem[{{Krumholz} {et~al.}(2009){Krumholz}, {McKee}, \&
  {Tumlinson}}]{kru09a}
---. 2009, \apj, 693, 216

\bibitem[{{Krumholz} \& {Tan}(2007)}]{kru07b}
{Krumholz}, M.~R. \& {Tan}, J.~C. 2007, \apj, 654, 304

\bibitem[{{Krumholz} \& {Thompson}(2007)}]{kru07}
{Krumholz}, M.~R. \& {Thompson}, T.~A. 2007, \apj, 669, 289

\bibitem[{{Lagos} {et~al.}(2012){Lagos}, {Bayet}, {Baugh}, {Lacey}, {Bell},
  {Fanidakis}, \& {Geach}}]{lag12}
{Lagos}, C.~d.~P., {Bayet}, E., {Baugh}, C.~M., {Lacey}, C.~G., {Bell}, T.,
  {Fanidakis}, N., \& {Geach}, J. 2012, arXiv/1204.0795

\bibitem[{{Larson}(1981)}]{lar81}
{Larson}, R.~B. 1981, \mnras, 194, 809

\bibitem[{{Leitherer} {et~al.}(1999)}]{lei99}
{Leitherer}, C. {et~al.} 1999, \apjs, 123, 3

\bibitem[{{Leroy} {et~al.}(2011){Leroy}, {Bolatto}, {Gordon}, {Sandstrom},
  {Gratier}, {Rosolowsky}, {Engelbracht}, {Mizuno}, {Corbelli}, {Fukui}, \&
  {Kawamura}}]{ler11}
{Leroy}, A.~K., {Bolatto}, A., {Gordon}, K., {Sandstrom}, K., {Gratier}, P.,
  {Rosolowsky}, E., {Engelbracht}, C.~W., {Mizuno}, N., {Corbelli}, E.,
  {Fukui}, Y., \& {Kawamura}, A. 2011, \apj, 737, 12

\bibitem[{{Lombardi} {et~al.}(2010){Lombardi}, {Alves}, \& {Lada}}]{lom10}
{Lombardi}, M., {Alves}, J., \& {Lada}, C.~J. 2010, \aap, 519, L7

\bibitem[{{Magdis} {et~al.}(2011){Magdis}, {Daddi}, {Elbaz}, {Sargent},
  {Dickinson}, {Dannerbauer}, {Aussel}, {Walter}, {Hwang}, {Charmandaris},
  {Hodge}, {Riechers}, {Rigopoulou}, {Carilli}, {Pannella}, {Mullaney},
  {Leiton}, \& {Scott}}]{mag11}
{Magdis}, G.~E., {Daddi}, E., {Elbaz}, D., {Sargent}, M., {Dickinson}, M.,
  {Dannerbauer}, H., {Aussel}, H., {Walter}, F., {Hwang}, H.~S.,
  {Charmandaris}, V., {Hodge}, J., {Riechers}, D., {Rigopoulou}, D., {Carilli},
  C., {Pannella}, M., {Mullaney}, J., {Leiton}, R., \& {Scott}, D. 2011, \apjl,
  740, L15

\bibitem[{{Mannucci} {et~al.}(2006){Mannucci}, {Della Valle}, \&
  {Panagia}}]{man06}
{Mannucci}, F., {Della Valle}, M., \& {Panagia}, N. 2006, \mnras, 370, 773

\bibitem[{{Meier} {et~al.}(2010){Meier}, {Turner}, {Beck}, {Gorjian}, {Tsai},
  \& {Van Dyk}}]{mei10}
{Meier}, D.~S., {Turner}, J.~L., {Beck}, S.~C., {Gorjian}, V., {Tsai}, C., \&
  {Van Dyk}, S.~D. 2010, \aj, 140, 1294

\bibitem[{{Mo} {et~al.}(1998){Mo}, {Mao}, \& {White}}]{mo98}
{Mo}, H.~J., {Mao}, S., \& {White}, S.~D.~M. 1998, \mnras, 295, 319

\bibitem[{{Narayanan}(2011)}]{nar11d}
{Narayanan}, D. 2011, arXiv/1112.1073

\bibitem[{{Narayanan} {et~al.}(2011{\natexlab{a}}){Narayanan}, {Cox},
  {Hayward}, \& {Hernquist}}]{nar11a}
{Narayanan}, D., {Cox}, T.~J., {Hayward}, C.~C., \& {Hernquist}, L.
  2011{\natexlab{a}}, \mnras, 412, 287

\bibitem[{{Narayanan} {et~al.}(2008){Narayanan}, {Cox}, {Kelly}, {Dav{\'e}},
  {Hernquist}, {Di Matteo}, {Hopkins}, {Kulesa}, {Robertson}, \&
  {Walker}}]{nar08a}
{Narayanan}, D., {Cox}, T.~J., {Kelly}, B., {Dav{\'e}}, R., {Hernquist}, L.,
  {Di Matteo}, T., {Hopkins}, P.~F., {Kulesa}, C., {Robertson}, B., \&
  {Walker}, C.~K. 2008, \apjs, 176, 331

\bibitem[{{Narayanan} {et~al.}(2011{\natexlab{b}}){Narayanan}, {Krumholz},
  {Ostriker}, \& {Hernquist}}]{nar11b}
{Narayanan}, D., {Krumholz}, M., {Ostriker}, E.~C., \& {Hernquist}, L.
  2011{\natexlab{b}}, \mnras, 418, 664

\bibitem[{{Narayanan} {et~al.}(2012){Narayanan}, {Krumholz}, {Ostriker}, \&
  {Hernquist}}]{nar12a}
{Narayanan}, D., {Krumholz}, M.~R., {Ostriker}, E.~C., \& {Hernquist}, L. 2012,
  \mnras, 421, 3127

\bibitem[{{Narayanan} {et~al.}(2006){Narayanan}, {Kulesa}, {Boss}, \&
  {Walker}}]{nar06b}
{Narayanan}, D., {Kulesa}, C.~A., {Boss}, A., \& {Walker}, C.~K. 2006, \apj,
  647, 1426

\bibitem[{{Oka} {et~al.}(1998){Oka}, {Hasegawa}, {Hayashi}, {Handa}, \&
  {Sakamoto}}]{oka98}
{Oka}, T., {Hasegawa}, T., {Hayashi}, M., {Handa}, T., \& {Sakamoto}, S. 1998,
  \apj, 493, 730

\bibitem[{{Pineda} {et~al.}(2008){Pineda}, {Caselli}, \& {Goodman}}]{pin08}
{Pineda}, J.~E., {Caselli}, P., \& {Goodman}, A.~A. 2008, \apj, 679, 481

\bibitem[{{Sandstrom} {et~al.}(2012)}]{san12}
{Sandstrom}, K.~M. {et~al.} 2012, arXiv/1212.1208

\bibitem[{{Sch{\"o}ier} {et~al.}(2005){Sch{\"o}ier}, {van der Tak}, {van
  Dishoeck}, \& {Black}}]{sch05}
{Sch{\"o}ier}, F.~L., {van der Tak}, F.~F.~S., {van Dishoeck}, E.~F., \&
  {Black}, J.~H. 2005, \aap, 432, 369

\bibitem[{{Schruba} {et~al.}(2012){Schruba}, {Leroy}, {Walter}, {Bigiel},
  {Brinks}, {de Blok}, {Kramer}, {Rosolowsky}, {Sandstrom}, {Schuster},
  {Usero}, {Weiss}, \& {Wiesemeyer}}]{sch12}
{Schruba}, A., {Leroy}, A.~K., {Walter}, F., {Bigiel}, F., {Brinks}, E., {de
  Blok}, W.~J.~G., {Kramer}, C., {Rosolowsky}, E., {Sandstrom}, K., {Schuster},
  K., {Usero}, A., {Weiss}, A., \& {Wiesemeyer}, H. 2012, arXiv/1203.4321

\bibitem[{{Shetty} {et~al.}(2011{\natexlab{a}})}]{she11a}
{Shetty}, R. {et~al.} 2011{\natexlab{a}}, \mnras, 412, 1686

\bibitem[{{Shetty} {et~al.}(2011{\natexlab{b}})}]{she11b}
---. 2011{\natexlab{b}}, \mnras, 415, 3253

\bibitem[{{Shirley} {et~al.}(2007){Shirley}, {Wu}, {Bussmann}, \&
  {Wootten}}]{shi07}
{Shirley}, Y.~L., {Wu}, J., {Bussmann}, R.~S., \& {Wootten}, A. 2007,
  arXiv/0711.4605, 711

\bibitem[{{Solomon} {et~al.}(1987){Solomon}, {Rivolo}, {Barrett}, \&
  {Yahil}}]{sol87}
{Solomon}, P.~M., {Rivolo}, A.~R., {Barrett}, J., \& {Yahil}, A. 1987, \apj,
  319, 730

\bibitem[{{Springel}(2005)}]{spr05b}
{Springel}, V. 2005, \mnras, 364, 1105

\bibitem[{{Springel} {et~al.}(2005){Springel}, {Di Matteo}, \&
  {Hernquist}}]{spr05a}
{Springel}, V., {Di Matteo}, T., \& {Hernquist}, L. 2005, \mnras, 361, 776

\bibitem[{{Strong} \& {Mattox}(1996)}]{str96}
{Strong}, A.~W. \& {Mattox}, J.~R. 1996, \aap, 308, L21

\bibitem[{{Strong} {et~al.}(2004){Strong}, {Moskalenko}, {Reimer}, {Digel}, \&
  {Diehl}}]{str04}
{Strong}, A.~W., {Moskalenko}, I.~V., {Reimer}, O., {Digel}, S., \& {Diehl}, R.
  2004, \aap, 422, L47

\bibitem[{{Tacconi} {et~al.}(2008)}]{tac08}
{Tacconi}, L.~J. {et~al.} 2008, \apj, 680, 246

\bibitem[{{Wilson}(1995)}]{wil95}
{Wilson}, C.~D. 1995, \apjl, 448, L97+

\bibitem[{{Wolfire} {et~al.}(2010){Wolfire}, {Hollenbach}, \& {McKee}}]{wol10}
{Wolfire}, M.~G., {Hollenbach}, D., \& {McKee}, C.~F. 2010, \apj, 716, 1191

\end{thebibliography}

\newpage

\end{document}